\documentclass[twocolumn]{aastex701}
\usepackage{multirow}
\usepackage{xcolor}

\begin{document}
\title{Optical Spectroscopy of a Candidate O-Star X-ray Binary in M33}

\author[0000-0003-3252-352X]{Margaret Lazzarini}
\affiliation{Department of Physics \& Astronomy, California State University Los Angeles, 5151 State University Drive, Los Angeles, CA 90032, USA}
\email[show]{mlazzar2@calstatela.edu}  

\author[0000-0001-8536-0547]{Lara Cullinane} 
\affiliation{Leibniz-Institut für Astrophysik Potsdam, An der Sternwarte 16, D-14482 Potsdam, Germany}
\email{lcullinane@aip.de}

\author[0000-0003-0394-8377]{Karoline Gilbert} 
\affiliation{Space Telescope Science Institute, 3700 San Martin Drive, Baltimore, MD 21218,USA}
\affiliation{The William H. Miller III Department of Physics \& Astronomy, Johns Hopkins University, 3400 N. Charles Street, Baltimore, MD 21218, USA}
\email{kgilbert@stsci.edu}

\author[0000-0001-8867-4234]{Puragra Guhathakurta} 
\affiliation{Department of Astronomy \& Astrophysics, University of California Santa Cruz, 1156 High Street, Santa Cruz, CA, USA}
\email{raja@ucolick.org}

\author[0000-0002-0206-1208]{Kyros Hinton}
\affiliation{Department of Astronomy, University of Washington, Box 351580, Seattle, WA 98195, USA}
\email{khinton001@gmail.com}

\author[0000-0003-2686-9241]{Daniel Stern} 
\affiliation{Jet Propulsion Laboratory, California Institute of Technology, 4800 Oak Grove Drive, Pasadena, CA 91009, USA}
\email{daniel.k.stern@jpl.nasa.gov}

\author[0000-0003-4122-7749]{O. Grace Telford} 
\affiliation{Department of Physics and Astronomy, University of Utah, 270 S 1400 E., Salt Lake City, UT 84112, USA}
\email{grace.telford@utah.edu}

\author[0000-0001-6320-2230]{Tobin M. Wainer}
\affiliation{Department of Astronomy, University of Washington, Box 351580, Seattle, WA 98195, USA}
\email{tobinw@uw.edu}

\author[0000-0002-7502-0597]{Benjamin F. Williams} 
\affiliation{Department of Astronomy, University of Washington, Box 351580, Seattle, WA 98195, USA}
\email{benw1@uw.edu}

\author[]{Robert Alexander} 
\affiliation{Department of Physics \& Astronomy, California State University Los Angeles, 5151 State University Drive, Los Angeles, CA 90032, USA}
\email{alexanderrobert@gmail.com}

\author[]{Ernesto Ramirez Jr.} 
\affiliation{Department of Physics \& Astronomy, California State University Los Angeles, 5151 State University Drive, Los Angeles, CA 90032, USA}
\email{ramirezernesto97@gmail.com}

\correspondingauthor{Margaret Lazzarini}

\begin{abstract}
We present new observations of a candidate O-type donor star in a high mass X-ray binary (HMXB) in M33. The candidate donor star is spatially coincident with a hard X-ray point source. The star's optical magnitude and colors are consistent with what is expected for a massive star. The four band optical/UV spectral energy distribution (SED) of the donor star is well fit by an O-giant model with an effective temperature of $\sim$43 kK, mass of 75$_{+9}^{-0.3}$ $M_{\odot}$, and surface gravity of log(g) of $\sim$3.8. We present four epochs of optical spectroscopy of the donor star taken with the Keck/DEIMOS spectrograph over the course of approximately one year. Features in the star's spectrum are consistent with an O-type star. The star's spectrum exhibits a strong H$\alpha$ emission line during all epochs of observation with a variable line profile. We measure the star's radial velocity for each epoch, which we use to fit for a systemic velocity. We find that the systemic velocity is offset from the local gas velocity from HI measurements, suggesting that the system has a peculiar velocity relative to the local gas. We observe a radial velocity shift in the He I absorption lines that is consistent with motion within a binary system.
\end{abstract}

\section{Introduction}

High mass X-ray binaries (HMXBs) are systems in which a black hole or neutron star accretes material from a massive stellar companion, or donor star. These systems are an important observable midpoint in the complex process of massive binary stellar evolution, which can result in the merger of a compact object binary, producing gravitational wave radiation \citep[e.g.,][]{Tauris2017,Kruckow2018}. HMXBs provide important constraints on theoretical models of massive  binary stellar evolution, serving as both individual test cases and population-level benchmarks that models must be able to reproduce.

Local Group galaxies present a unique opportunity to observe galaxy-wide populations of HMXBs. These galaxies are close enough to resolve into their constituent stars with optical surveys \citep[e.g.,][]{Dalcanton,Williams2021}. At X-ray wavelengths, surveys can identify systems with luminosities as low as $\sim$a few $\times$10$^{35}$ erg s$^{-1}$ \citep{Tullmann2011,Williams2015,Williams2018}. Unlike in more distant galaxies, where only the most luminous systems can be observed, this limit enables characterization of the HMXB population over a much broader luminosity range. Population scale properties of HMXBs including HMXB production rates, fractions of O versus B type donor stars, and dominant ages of individual systems can then be compared with galaxy-scale properties including star formation rate (SFR), stellar mass, and metallicity to directly tie the local stellar environment to HMXB formation and evolution \citep[e.g.,][]{Antoniou2010,Antoniou&Zezas2016,Garofali2018,Lazzarini2018,Antoniou2019,Lazzarini2021,Lazzarini2023}.

Despite the value of these systems and the proximity of M33, only one HMXB system has been spectroscopically confirmed in this galaxy: M33 X-7 \citep{Orosz2007,Ramachandran2022}. In fact, it is the only spectroscopically confirmed HMXB in both M31 and M33, the nearest spiral galaxies to the Milky Way. Previous studies have used overlapping surveys by the Hubble Space Telescope (HST) and the Chandra X-ray Observatory to identify HMXB candidates in both of these galaxies \citep{Garofali2018,Lazzarini2018,Lazzarini2021,Lazzarini2023}. These studies identified 40 and 65 HMXB candidates in M31 and M33, respectively, as hard X-ray point sources spatially coincident with optical counterparts that were point sources with colors and magnitudes consistent with being a massive star. To spectroscopically confirm these as HMXB systems, follow up time-resolved optical spectroscopy is needed to confirm the spectral type of the candidate donor star and constrain the mass of the compact object.

In this paper we present optical spectroscopy of an HMXB candidate in M33 identified through comparison of Chandra and HST imaging data by \citet{Lazzarini2023}. We obtained four epochs of observations over 11 months with the Keck/DEIMOS spectrograph to characterize this candidate donor star in more detail. While the sparse time sampling of these data is not sufficient to fully constrain the orbital elements of the system, with these data we can confirm if the candidate donor is a high mass star, and can measure any velocity shifts that would indicate the presence of an unseen binary companion.

In Section \ref{sec:data} we present an overview of the HST photometry used to identify the HMXB candidate and the new optical spectroscopy presented in this paper. In Section \ref{sec:results} we present an overview of SED fitting used to determine the physical properties of the candidate donor star, spectral classification of the candidate donor star using the optical spectrum, measurements of the candidate donor star's radial velocity, the time evolution of the line profiles of the H$\alpha$ and H$\beta$ lines, and constraints on the orbital period from the radial velocity (RV) measurements. In Section \ref{sec:discussion} we discuss the implications of these results and in Section \ref{sec:conclusions} we summarize our conclusions.

\section{Data}\label{sec:data}
%The source's optical properties as observed with HST imaging and Keck/DEIMOS optical spectroscopy are discussed in this paper. Each is discussed in more detail below.

\subsection{HST Photometry}
\begin{figure*}
\centering
\includegraphics[width=0.9\textwidth]{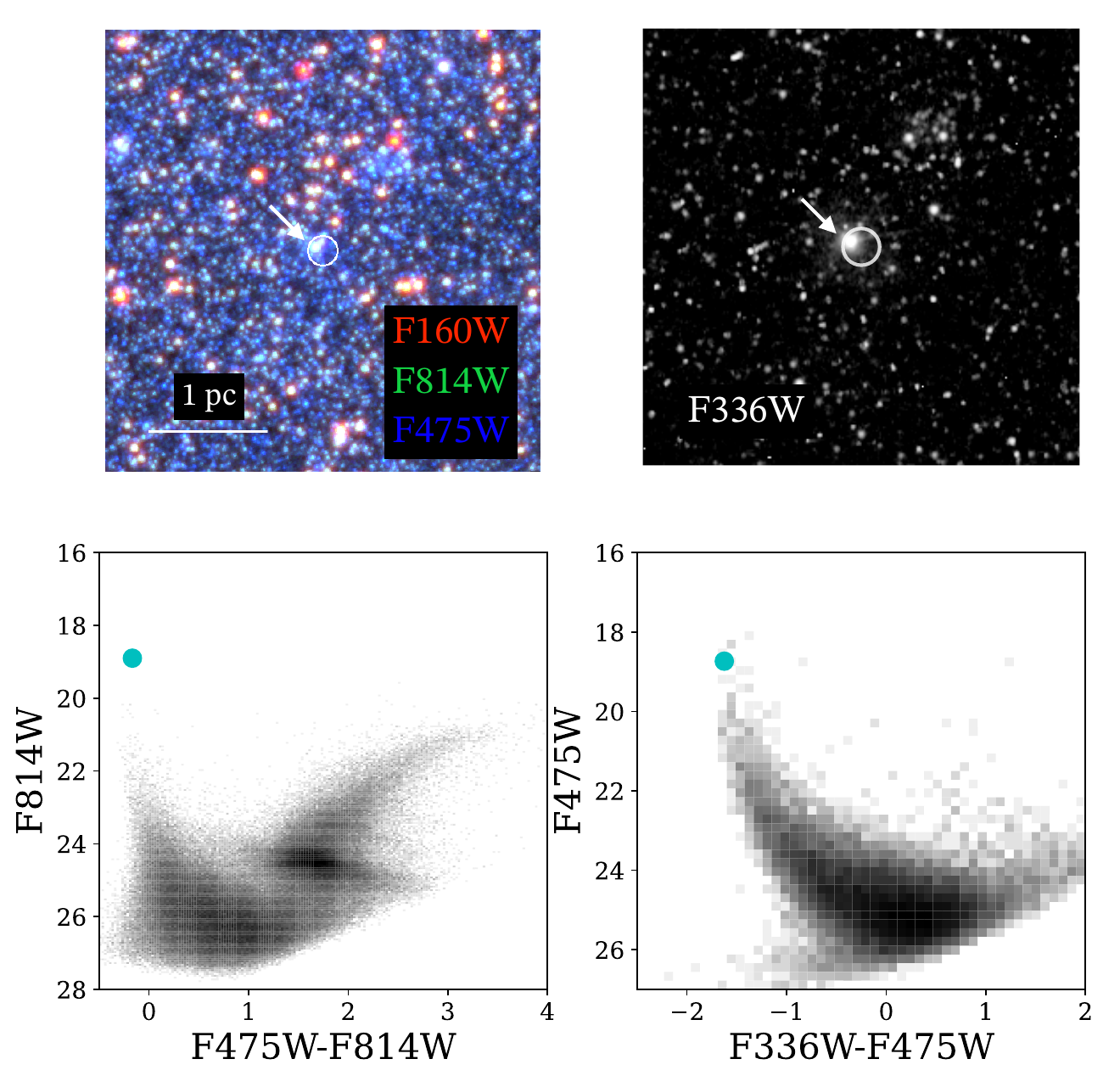}
\caption{Images and Hess diagrams indicating the candidate donor star. \textbf{Upper left: }A RGB optical image created with mosaic images from the PHATTER survey with a 1 parsec scale bar in the lower left corner. The X-ray source's 1$\sigma$ positional error from \citet{Tullmann2011} is plotted with a white circle. A white arrow points to the star. The RGB optical image was created using the F160W (red), F814W (green), and F475W (blue) bands. \textbf{Upper right: }Single band UV image in the F336W band. The X-ray source's 1$\sigma$ positional error is plotted with a white circle. A white arrow points to the star. \textbf{Lower left: }Optical Hess diagram showing all stars in the PHATTER survey that pass photometric quality cuts from \citet{Williams2021}. The cyan circle marks the position of this star on the Hess diagram. \textbf{Lower right:} Optical/UV Hess diagram showing all stars in the PHATTER survey that pass photometric quality cuts. The cyan circle marks the position of the star on the Hess diagram.}
\label{fig:finders_cmds}
\end{figure*}

The optical counterpart discussed in this paper was identified by \citet{Lazzarini2023} using HST imaging and photometry from the Panchromatic Hubble Andromeda Treasury: Triangulum Extended Region (PHATTER) survey, an HST survey that covered the inner disk of M33 in six photometric bands spanning near-IR to near-UV wavelengths, producing a photometric catalog of over 20 million individual stars \citep{Williams2021}. \citet{Lazzarini2023} identified as optical counterparts to X-ray point sources in the Chandra ACIS Survey of M33 \citep[ChASeM33;][]{Tullmann2011} and compiled a sample of HMXB candidates. Priority HMXB candidates were identified using both X-ray and optical properties. The X-ray point source needed to have hardness ratios consistent with an accreting compact object. There also needed to be an optical counterpart within the X-ray source's 1-$\sigma$ positional error with optical/UV colors and magnitudes consistent with being a massive star at the distance of M33. The optical counterpart, which is the subject of this paper, was identified in that paper as 013402.8833+304151.341, which is the source's name (and position) in the PHATTER survey \citep{Williams2021}. The source's magnitude in the HST filters from the PHATTER survey are listed in Table \ref{table: phatter_photometry}. We present optical and UV images of the $\sim$10$^{\prime \prime}\times$10$^{\prime \prime}$ region around the star from the PHATTER mosaic imaging and Hess diagrams indicating the star's color and magnitude relative to other stars observed in the PHATTER survey in Figure \ref{fig:finders_cmds}.

\begin{deluxetable*}{ccccccc}
\tablecaption{HST photometry for HMXB companion star candidate\label{table: phatter_photometry}}
\tablehead{
\colhead{PHATTER Catalog Name} &  
\colhead{F275W} &
\colhead{F336W} & 
\colhead{F475W} & 
\colhead{F814W} &
\colhead{F110W} &
\colhead{F160W} 
}
\startdata
013402.8833+304151.341 & 16.67$\pm$0.016 & 17.11$\pm$0.003 & 18.73$\pm$0.001 & 18.90$\pm$0.002 & 19.08$\pm$0.002 & 19.12$\pm$0.003 \\
\enddata
\tablecomments{List of Vega magnitudes for the candidate HMXB companion star discussed in this paper from the PHATTER survey \citep{Williams2021}.}
\end{deluxetable*}

\subsection{Optical Spectroscopy}
We obtained four epochs of optical spectroscopy for this source with the 10-m Keck II telescope at the W.M. Keck Observatory, located on the summit of Mauna Kea, Hawaii (USA). Observations were carried out during the fall of 2022 and 2023. All spectroscopic data come from observations taken with the DEIMOS Spectrograph \citep{Faber2003}. The slitmasks including the star of interest in this paper were part of a larger spectroscopic follow-up effort to obtain optical spectroscopy for candidate HMXB companion stars identified in M31 and M33 by \citet{Lazzarini2021,Lazzarini2023} that will be presented in future work.

All observations were carried out with the same spectrographic setup. Slitmasks were observed using the 600ZD grating, which has a resolving power of $R\sim$2000, and a central wavelength of 7200 \AA. The GG455 long-pass spectroscopy order blocking filter was used. The shortest wavelength covered in the 2022-11-18 observation is $\sim$5000~\AA~ compared to $\sim$4500~\AA~ in each of the 2023 observations. A different slitmask was used for the 2022 versus 2023 observations and the star's position on the mask resulted in slightly different wavelength coverage.

Each observation had an exposure time of approximately one hour, and each one hour observation was divided into three $\sim$20 minute exposures. A summary of observations of the HMXB companion star candidate discussed in this paper are listed in Table \ref{table: deimos_observations}. 

\begin{deluxetable}{cccc}
\tablecaption{Summary of DEIMOS observations of the source\label{table: deimos_observations}}
\tablehead{
\colhead{Date [UTC]} &  
\colhead{Start Time [UTC]} & 
\colhead{Airmass} & 
\colhead{Exp. Time [s]}
}
\startdata
2022-11-18 & 04:48:56.29 & 1.38 & 1200 \\
2022-11-18 & 05:20:02.14 & 1.29 & 1200 \\
2022-11-18 & 05:41:10.01 & 1.22 & 1200 \\
2023-08-13 & 12:07:47.40 & 1.2 & 1100 \\
2023-08-13 & 12:27:12.76 & 1.15 & 1100 \\
2023-08-13 & 12:46:42.19 & 1.11 & 1100 \\
2023-09-10 & 11:49:47.53 & 1.05 & 1000 \\
2023-09-10 & 12:07:34.47 & 1.04 & 1000 \\
2023-09-10 & 12:25:19.50 & 1.02 & 1000 \\
2023-10-10 & 11:43:06.11 & 1.05 & 1200 \\
2023-10-10 & 12:04:11.51 & 1.07 & 1200 \\
2023-10-10 & 12:25:20.48 & 1.11 & 1200
\enddata
\tablecomments{Summary of Keck/DEIMOS observations of the HMXB companion star candidate discussed in this paper. Exposures for each night were combined together to produce the spectra used to identify absorption/emission lines and measure radial velocities.}
\end{deluxetable}

The DEIMOS spectra were reduced using the \verb|PypeIt| software package \citep{pypeit:joss_arXiv,pypeit:joss_pub,pypeit:zenodo}. Reductions were performed using the standard workflow for Keck/DEIMOS spectra presented in the \verb|PypeIt| documentation. The resulting one-dimensional spectra were flux calibrated using the \verb|PypeIt| provided sensitivity function and all spectral data analyzed in this paper were continuum-normalized. \verb|PypeIt| applies heliocentric corrections to each spectrum. The \verb|PypeIt| \verb|pypeit_coadd_1dspec| command was used to coadd one-dimensional spectra.

\section{Results}\label{sec:results}
The HMXB companion star candidate was characterized using several methods in this study and in previous work. The star's properties were measured using spectral energy distribution (SED) fitting by \citet{Lazzarini2023}, the results of which are summarized in Section \ref{sec: sed_fitting}. We used the star's optical spectrum to identify absorption and emission features to confirm the star's spectral type. We measured radial velocities in each epoch. We analyze the evolution of the line profile of the Balmer lines over the course of our observations. Lastly, we used \texttt{The Joker} \citep{Price-Whelan2017}, a Monte Carlo sampler designed to fit orbital parameters for sparse radial velocity datasets, to look for a periodic signal in our RV measurements to constrain the binary orbit. We discuss each of these results in detail in the following sub-sections.

\subsection{Stellar Properties from SED fitting}\label{sec: sed_fitting}
\citet{Lazzarini2023} used four photometric bands: F275W, F336W, F475W, and F814W, which span near-ultraviolet and optical wavelengths, to perform SED fitting using the Bayesian Extinction and Stellar Tool \citep[\texttt{BEAST};][]{Gordon2016}, which fits observed SEDs with stellar evolution models using Bayesian methods. The \texttt{BEAST} fits for stellar parameters including age, mass, metallicity, distance and parameters describing dust along the line of sight including dust column density ($A_{V}$), average grain size ($R_{V}$), and $f_{A}$, which describes the distribution of different types of dust grains within the Local Group. Stellar parameters derived from the initial fit include luminosity, effective temperature, radius, and surface gravity. We present the best fit stellar parameters for our HMXB companion star candidate in Table \ref{table: BEAST_parameters}. The star was classified as an O-type giant using the ranges of effective temperature, luminosity, and radius presented in \citet{Lamers&Levesque2017}.

\begin{deluxetable}{cc}
\tablecaption{Stellar properties measured via SED fitting\label{table: BEAST_parameters}}
\tablehead{
\colhead{Property} &  
\colhead{SED fit value} 
}
\startdata
log($T_{\rm eff})$ [K] & 4.6$^{+0.1}_{-0.1}$ \\
log($L$) [erg s$^{-1}$] & 6.0$^{+0.2}_{-0.1}$\\
$R$ [$R_{\odot}$] & 17.9$^{+0.5}_{-1.7}$ \\
$A_{V}$ [mag] &  0.4$^{+0.1}_{-0.1}$\\
current $M$ [$M_{\odot}$] & 75$^{+1}_{-9}$\\
\enddata
\tablecomments{Summary of best-fit stellar parameters derived from SED fitting of the star's PHATTER photometry with the \texttt{BEAST} \citep{Gordon2016}. SED fit values listed are the best values returned by the \texttt{BEAST} and the error bars represent the 16th and 84th percentile values for each parameter. For a more detailed discussion of how the best fit value and errors were measured, see Section 3.2 of \citet{Lazzarini2023}.}
\end{deluxetable}

\subsection{Spectral Classification using Keck/DEIMOS spectra}
\begin{figure*}
\centering
\includegraphics[width=\textwidth]{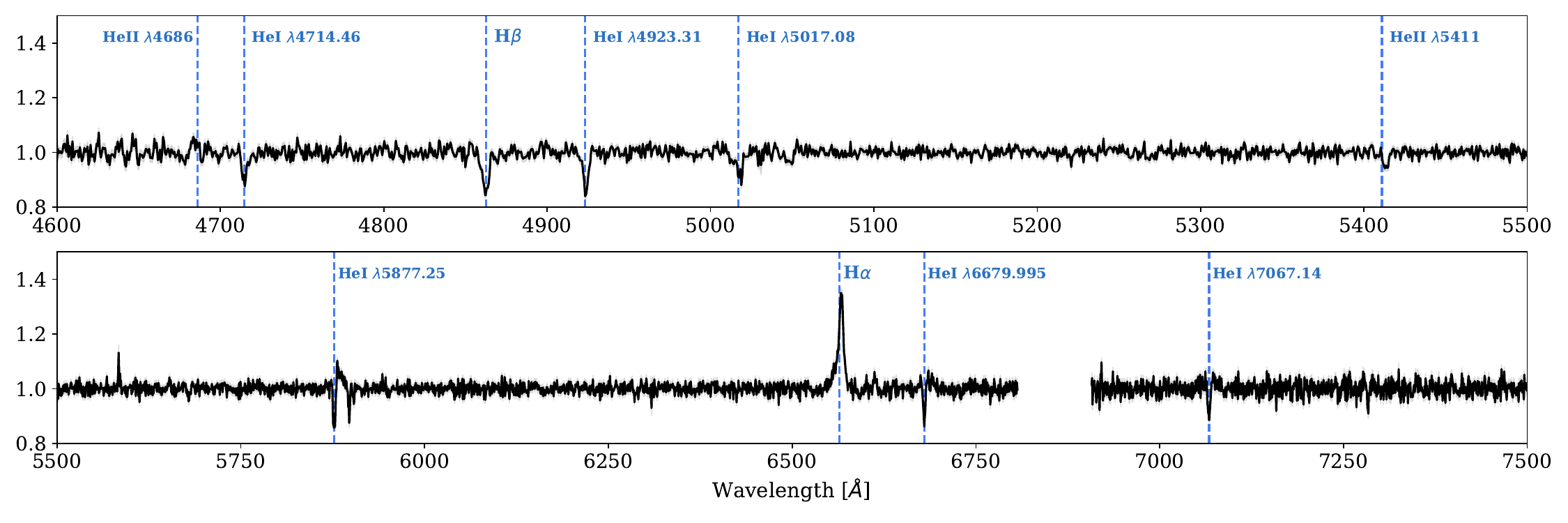}
\caption{Combined spectrum for our candidate O-type donor star. The spectrum was produced by co-adding three epochs of observation from 2023 and is shown in the rest frame of the systemic velocity measured in Section \ref{sec:system_velocity}. The spectrum has been continuum normalized. The y-axis shows relative flux. The star's spectrum exhibits both He I and He II absorption features, confirming that it is an O-type star. The wavelengths of the indicated spectral features are listed as rest wavelengths in a vacuum.}
\label{fig:full_spectrum}
\end{figure*}

We confirm that the candidate donor star's spectrum is consistent with being an O-type star. We visually inspected the combined spectrum with data from all three 2023 epochs of observation to increase signal to noise. We present the combined spectrum in Figure \ref{fig:full_spectrum}, which combines the three spectra from the observations in August 2023, September 2023, and October 2023 in observed wavelength. The spectrum has been continuum-normalized and the y-axis shows flux relative to the continuum. The spectrum shows a strong H$\alpha$ emission line, H$\beta$ in absorption, multiple He I absorption lines and a He II $\lambda$5411~\AA~ absorption line. The He II 5411~\AA~ line is fairly weak in the single epoch spectra, but is visible in the combined spectrum shown in Figure \ref{fig:full_spectrum}. The He II $\lambda$4684~\AA~ line also appears to be present, partially filled in by emission. This collection of spectral features is consistent with a late type O-star due to the weak He II $\lambda$5411~\AA~ absorption compared to the He I absorption lines. Due to the wavelength range in our DEIMOS spectra ($\lambda > 4500 \AA$) we cannot compute the line ratios needed for a more granular spectral classification, such as those discussed in \citet{Martins2018}. 

It is interesting to note that there is some tension between the implied mass based on the star's spectrum and its SED fit. The late O-type classification from the star's optical spectrum, while not precise, implies that the star likely has a mass of less than the current mass of 75 $M_{\odot}$ measured via SED fitting, which would be more consistent with an early type O star. It is our conjecture that the SED fit may have overestimated the star's mass due to UV contamination from the compact object's accretion disk, which may have made the star appear brighter in the UV than it is. Our spectra for the star do not extend blueward of $\sim$4500~\AA~, which makes precise spectral typing for an O-type star challenging because the line ratios typically used to determine the spectral types of massive stars include lines not covered in our observations \citep[e.g.,][]{Sota2011}. UV or bluer optical spectroscopy will be required to make a more firm measurement of the star's precise temperature and mass. However, these spectra clearly identify it as a massive O star in M33.

\subsection{Radial Velocity Measurements}\label{sec:rv_measurements}
We measured the star's radial velocity using the He I absorption lines, as these are assumed to come from the star's surface, providing the best estimate of its movement. We did not measure radial velocities using the H$\alpha$ or H$\beta$ lines, as these lines are known to be affected by stellar winds and accretion onto the compact object and are generally not reliable for radial velocity measurements, especially for a massive star in an HMXB. We discuss the Balmer lines in more detail in Section \ref{sec:balmer_lines}. 

We measure the star's radial velocity at each epoch of observation using the He I $\lambda$6679~\AA~ line. We used a series of quality checks to determine which lines would provide reliable radial velocity measurements. The He I $\lambda$6679~\AA~ line was the only line in absorption, and thus assumed to be coming from the star's photosphere, that passed these quality checks in all four epochs. 

The three quality checks include a measurement of the local continuum signal to noise ratio (SNR), the line's significance relative to the continuum, and the robustness of the line's fit. We measured the continuum SNR within 100~\AA~ of the line's rest wavelength. We only performed radial velocity measurements on lines with local continuum SNR $>$12. We measured each line's significance relative to the local continuum and only performed radial velocity measurements on lines detected at $\geq$4$\sigma$ significance, defined as the line's contrast relative to the continuum divided by the local continuum RMS. For a given line, we defined a fit window around the line's detected central wavelength and used spectral data within this window to perform a fit to the line. This ensured that the choice of fit window did not impact the best fit central wavelength for the line. We ran our fit for each line five times, varying the center of the fit window by [-2,-1,0,1,2] pixels and measured the best fit line center in each shifted window. We only accepted lines for which the change in the best-fit central wavelength was much less than the 1$\sigma$ error on the line's central wavelength from the fit.

We measured the line's radial velocity by fitting the line's observed central wavelength relative to the line's rest wavelength in a vacuum. We fit each line individually using the python tool \verb|lmfit| \citep{lmfit}. The \verb|lmfit| tool is a non-linear least-squares minimization and curve-fitting tool written in python with built in models ideal for fitting stellar spectral features. We fit the spectral lines in the star's spectrum using a Voigt profile. In the fitting process, the inverse of the 1$\sigma$ errors on the star's flux measurement were used to weight the flux values. 

\begin{figure}
\centering
\includegraphics[width=0.45\textwidth]{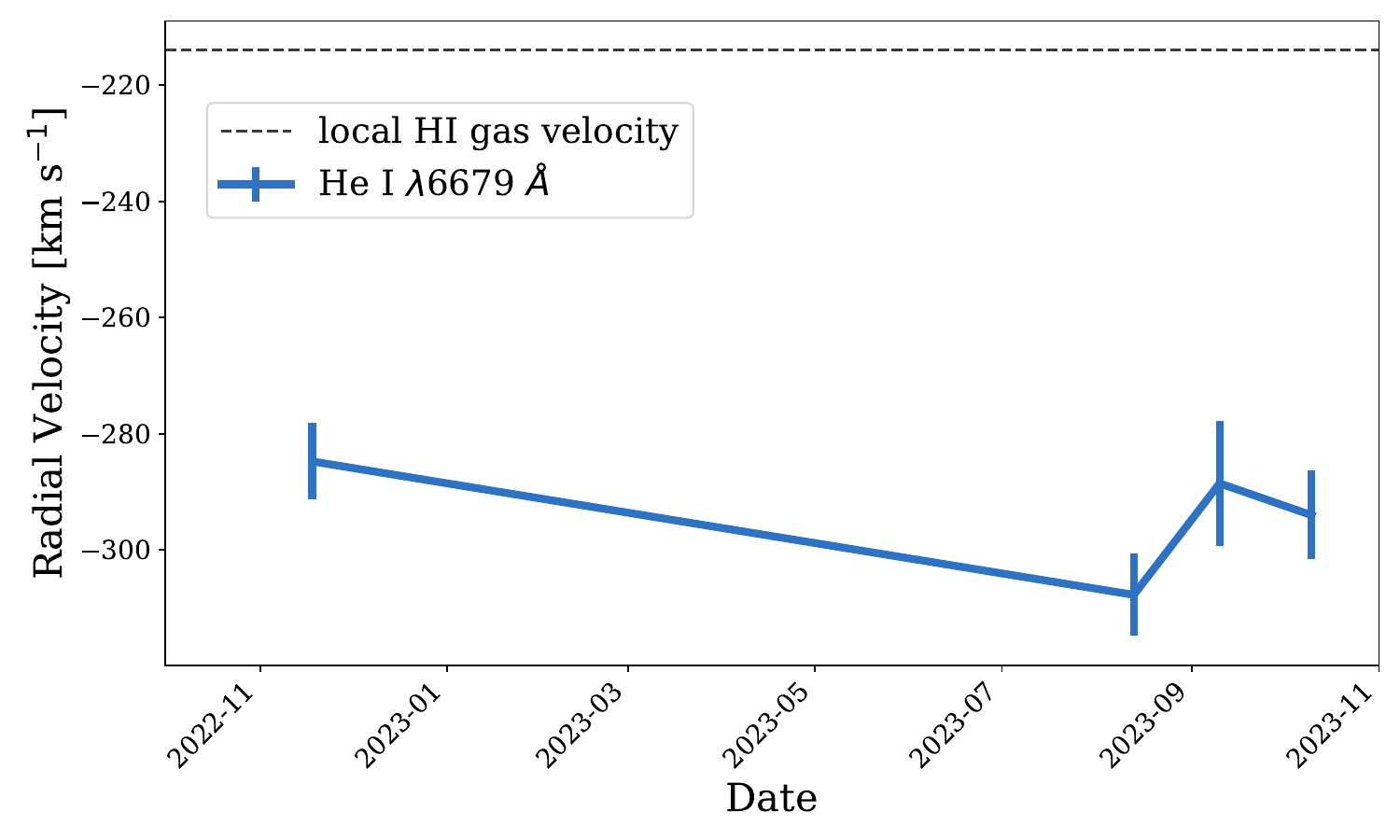}
\caption{The star's radial velocity for each epoch of observation measured with the He I $\lambda$6679.995~\AA~ line. The horizontal dashed line shows the local HI gas velocity measured from HI maps \citep{Koch2018}.}
\label{fig:rv}
\end{figure}

We report radial velocities for the He I $\lambda$6679~\AA~ line in Table \ref{table: rvs}. These radial velocities do not have the motion of M33 relative to the Milky Way or motion within the disk of M33 removed, although the standard \verb|PypeIt| data reduction applies a heliocentric correction. We plot the star's radial velocity measured for each of the four epochs in Figure \ref{fig:rv}. We include a horizontal line showing the local gas velocity measured from HI maps at the location of the star \citep{Koch2018} for reference. To investigate the statistical significance of the observed RV shift, we fit the data to a constant RV ($\chi^{2}=5.97$, $p=0.11$). This fit indicates that the four epochs show suggestive, though not definitive, evidence for a radial velocity shift. Additional observations will be needed to better constrain the significance and physical interpretation of the systems radial velocity variability.

The binary system also appears to be moving relative to the local gas based on the offset from the local HI gas velocity. Current fits (discussed in more detail in Section \ref{sec:the_joker}) suggest that this system is moving relative to its local environment, which is not an unexpected outcome for an HMXB, since they can be imparted a peculiar velocity via supernova kick when the primary becomes a compact object. Another possible explanation is that this offset is due to motion within the binary system. Additional epochs of observation with higher time resolution are required to definitely determine the cause of the velocity offset. More discussion of the system's velocity relative to the local gas is presented in Section \ref{sec:system_velocity}.

\begin{deluxetable*}{cccc}
\tablecaption{Radial velocity measurements and line diagnostics for He I $\lambda$6679.995~\AA~\label{table: rvs}}
\tablehead{  
\colhead{Date (UTC)} & 
\colhead{Velocity [km s$^{-1}$]} & 
\colhead{Continuum SNR} &
\colhead{Line Significance}
}
\startdata
2022-11-18 & $-284.8\pm6.6$ & 12.6 & 7.1\\
2023-08-13 & $-307.7\pm7.1$ & 41.0 & 7.9\\
2023-09-10 & $-288.5\pm10.8$ & 22.1 & 4.0\\
2023-10-10 & $-294.0\pm7.6$ & 20.4 & 4.8\\
\enddata
\tablecomments{Radial velocities for the candidate donor star measured with the He I $\lambda$6679.995~\AA~ line as described in Section \ref{sec:rv_measurements}. We include the local continuum SNR and the line's significance, which were used to determine which lines had sufficient data quality for a robust RV measurement.}
\end{deluxetable*}

We attempted to further constrain the system's velocity by searching for a counterpart in the ZTF survey, limiting magnitude of $g\sim$20.5 \citep{vanRoestel2021}. However, in the complex background of M33, isolating a star at the resolution of ZTF is difficult. Nevertheless, we searched for the star's counterpart in ZTF's data release 23 \citep{ztf}. The purpose of searching in the ZTF data was to detect a period of optical variability. While there were two detections in the ZTF catalog within 0.2$^{\prime \prime}$ of our source position with similar magnitudes, neither had a sufficient number of detections to constrain the orbital period.

\subsection{Evolution of Hydrogen Balmer Lines}\label{sec:balmer_lines}
\begin{figure*}
\centering
\includegraphics[width=0.8\textwidth]{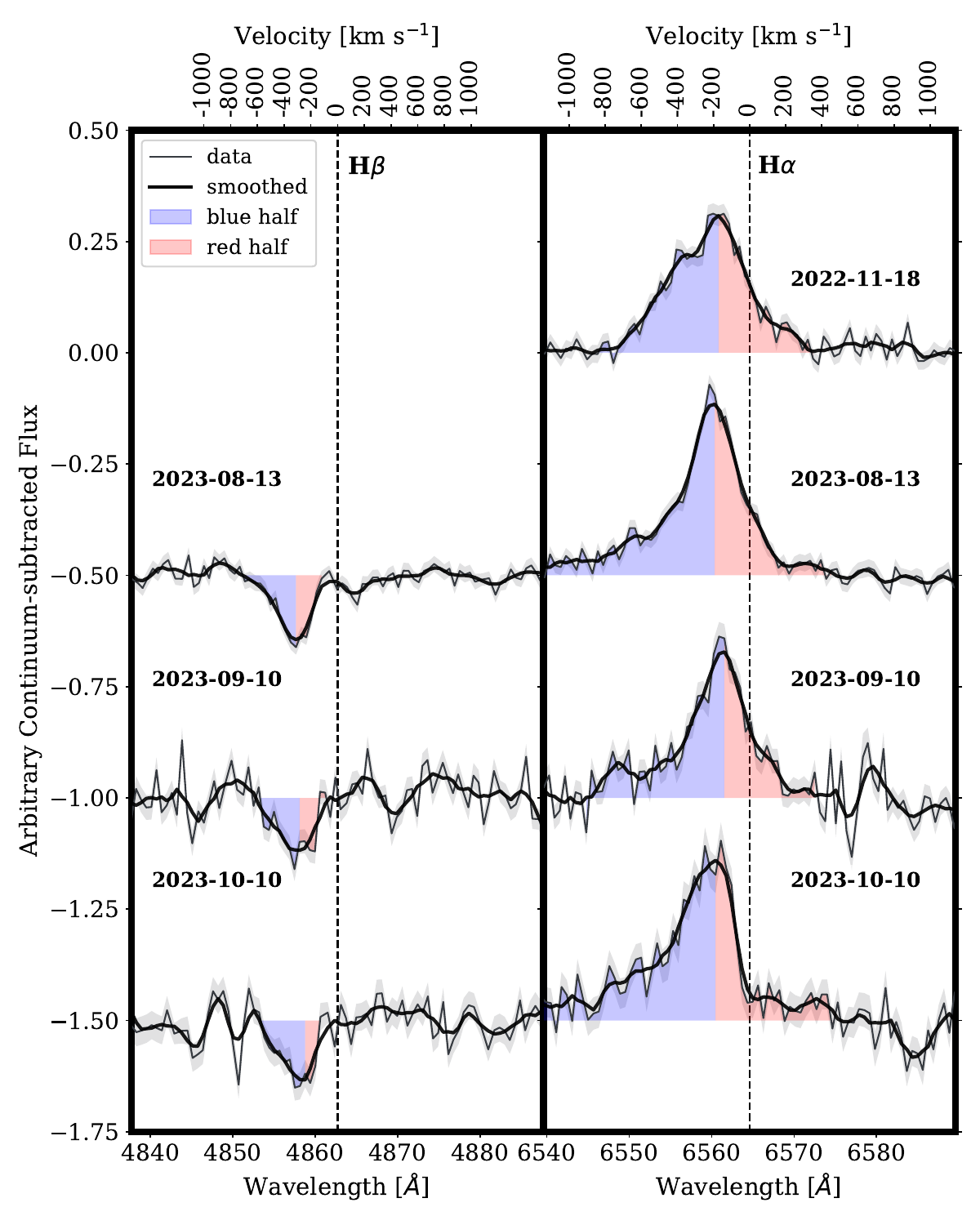}
\caption{The line profile of the H$\alpha$ and H$\beta$ lines over the course of our observations. An asymmetrical blue wing emerges in the H$\alpha$ emission line in the September and October 2023 observations. The H$\beta$ line profile does not change significantly, but a feature emerges blue-ward of the H$\beta$ absorption line in parallel with the shift in the H$\alpha$ emission line profile changes. The spectra are plotted in observed wavelength, they have not been shifted to match the radial velocity measured for each epoch. The zero velocity (dashed vertical) line is the rest velocity of H$\alpha$ and H$\beta$. The shading around the solid black line indicates the 1$\sigma$ uncertainty on the star's flux measurements.}
\label{fig:balmer_line_evolution}
\end{figure*}

The H$\alpha$ and H$\beta$ lines in the star's spectrum exhibit line profile evolution across the four epochs of observation, as shown in Figure \ref{fig:balmer_line_evolution}. Visual inspection of the 2-dimensional spectrum near H$\alpha$ shows no obvious spatially extended emission perpendicular to the stellar trace, suggesting no clear contamination from nebular emission.

The H$\alpha$ line is in emission during all four epochs, and the shape of the line profile shifts over the course of the four epochs. The line is asymmetrical during all observations, with an enhanced blue wing compared to the red wing. To quantify the evolution of the H$\alpha$ line profile, we measure the line's equivalent width (EW) as a measure of the strength of emission, the line's full width at half maximum (FWHM) as a measure of the velocity spread of the H$\alpha$ emitting gas, and the line's asymmetry. The H$\beta$ line is in absorption during the three 2023 epochs during which it was observed; the wavelength of H$\beta$ was not covered in the November 2022 observations due to the source's position on the slitmask, as described in Section \ref{sec:data}. The H$\beta$ line also shows asymmetry in its shape with an enhanced blue wing. A feature emerges blue-ward of the H$\beta$ absorption line in parallel with the shift in the H$\alpha$ line profile changes, as shown in Figure \ref{fig:balmer_line_evolution}. Higher resolution spectroscopic data are needed to definitively constrain the origin of the evolution of both the H$\alpha$ and H$\beta$ line profiles.

The EWs of H$\alpha$ and H$\beta$ lines can be used to determine if an HMXB system has a supergiant or Be donor star \citep{Fabregat1996,Reig1996}. We measure the EW of the H$\alpha$ and H$\beta$ lines for the star at each epoch of observation which we list in Table \ref{table: ews}. We measure the EW using the continuum-normalized spectrum and set the bounds within which the EW is calculated by visually inspecting where the wings of the H$\alpha$ emission line end. In Figure 2 of \citet{Reig1996}, they plot the H$\beta$ EW versus H$\alpha$ EW for OB supergiant stars versus Be stars with Balmer emission from the literature. The Balmer line EWs measured for the star in this paper lies in the giant/supergiant region of the diagram. This is consistent with our SED fits implying that the star is likely an O-giant and suggesting that this HMXB is likely a wind-fed system with an O-giant donor star. 

\begin{deluxetable}{cccc}
\tablecaption{Equivalent widths for H$\alpha$ and H$\beta$ lines, and resulting ratio\label{table: ews}}
\tablehead{
\colhead{Obs. Date (UTC)} &  
\colhead{H$\alpha$ EW [$\AA$]} & 
\colhead{H$\beta$ EW [$\AA$]} & 
\colhead{H$\alpha$/H$\beta$ EW Ratio}
}
\startdata
2022-11-18 & -3.54 & -- & --\\
2023-08-13 & -3.82 & 0.71 & -5.38\\
2023-09-10 & -3.08 & 0.62 & -4.97\\
2023-10-10 & -3.68 & 0.63 & -5.84\\
\enddata
\tablecomments{Equivalent widths of the H$\alpha$ and H$\beta$ lines and their ratio measured during each epoch of observation. There is no measurement of the H$\beta$ equivalent width on 2022-11-18 because the spectrum did not extend to short enough wavelengths to measure it. Negative EWs are in emission and positive EWs are in absorption.}
\end{deluxetable}

\begin{figure}
\centering
\includegraphics[width=0.45\textwidth]{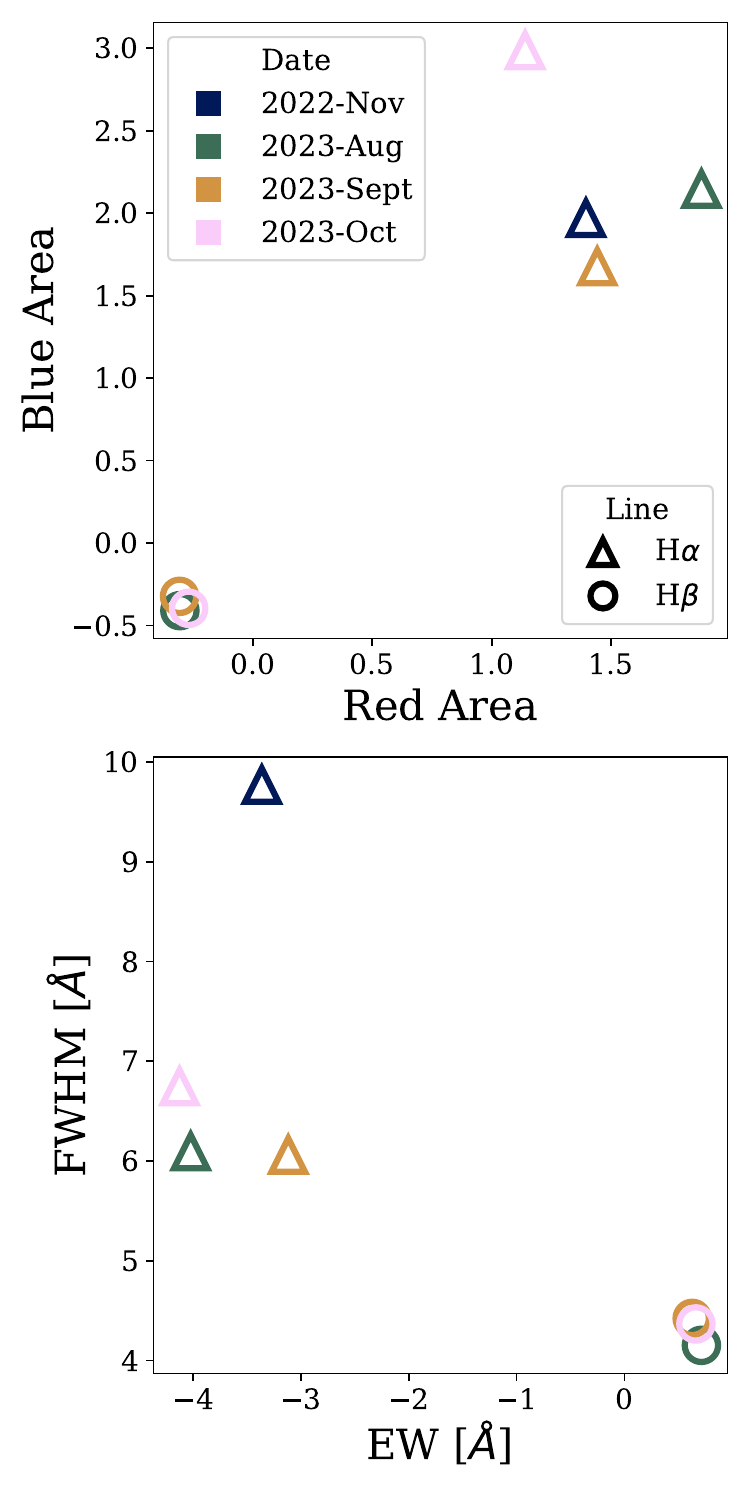}
\caption{Measurements of the H$\alpha$ (triangles) and H$\beta$ (circles) lines used to quantify their line profile evolution. \textbf{Upper panel:} Areas of the red half and blue half of the Balmer lines during each epoch of observation, as shown in Figure \ref{fig:balmer_line_evolution}. Points plotted with circles indicate measurements for H$\beta$ and points plotted with triangles indicate measurements for H$\alpha$. \textbf{Lower panel:} EW versus FWHM for the H$\alpha$ and H$\beta$ lines.}
\label{fig:balmer_line_diagnostics}
\end{figure}

To quantify the asymmetry of the H$\alpha$ line profile, we compare the area of the blue and red halves of the emission line, which requires defining the ``line center''. Due to the line's asymmetry, fitting a Voigt profile or even a skewed Voigt profile did not properly identify the line peak. Thus we used the peak of the line profile after applying smoothing, convolving the raw spectrum with a 3 pixel Gaussian. Using the smoothed data to find the peak avoided issues with noise. We plot the area of the red and blue halves of each line in Figure \ref{fig:balmer_line_diagnostics}.

It is challenging to measure a radial velocity for the Balmer lines due to their shifting line profile and asymmetric shape. Instead of attempting to measure a radial velocity, we measure the velocity at the ``line center'' that was used to separate the blue and red halves of the line for the investigation of the evolution of the line's asymmetry. We use this as a proxy for the line's velocity, which can hint at its physical origin within the system. 

We list the approximate centroid velocity we measure for each line, rounded to the nearest integer, in Table \ref{table: balmer_velocities}. The H$\beta$ centroid velocities are very similar to the radial velocities measured with the He I lines, suggesting that the H$\beta$ line also originates from the donor star's surface. The H$\alpha$ line centroid velocities are significantly offset from the He I and H$\beta$ velocities, suggesting a different origin within the system. The H$\alpha$ emission line is also quite broad compared to the H$\beta$ and He I $\lambda$6679~\AA~ absorption lines. The blue wing of the H$\alpha$ line extends to velocities of approximately 800$-$1000 km s$^{-1}$ and the red wing extends to velocities of about 400 km s$^{-1}$, as shown in Figure \ref{fig:balmer_line_evolution}. %XXX Update to velocity relative to the line centroid?

\begin{deluxetable}{ccc}
\tablecaption{Velocities measured for H$\alpha$ and H$\beta$ lines at the line centroids \label{table: balmer_velocities}}
\tablehead{
\colhead{Observation date (UTC)} &  
\colhead{H$\alpha$ [km s$^{-1}$]} & 
\colhead{H$\beta$ [km s$^{-1}$]} 
}
\startdata
2022-11-18 & -173 & -- \\
2023-08-13 & -223 & -309 \\
2023-09-10 & -170 & -320 \\
2023-10-10 & -160 & -279 \\
\enddata
\tablecomments{Heliocentric-corrected velocities measured at the peak/trough of the H$\alpha$ and H$\beta$ lines during each epoch of observation. There is no measurement of the H$\beta$ velocity on 2022-11-18 because the spectrum did not extend to short enough wavelengths to measure it. We determined the peak (for H$\alpha$ lines in emission) or trough (for H$\beta$ lines in absorption) by smoothing the continuum-normalized spectrum. We do not report these as stellar photospheric radial velocities, but we compared the Balmer line velocities with the star's radial velocity measured with He I lines as an indication for the physical origin of the H$\alpha$ and H$\beta$ lines as described in Section \ref{sec:balmer_lines}.}
\end{deluxetable}

\subsection{Constraining Orbital Period with Radial Velocity Measurements}\label{sec:the_joker}
We used \texttt{The Joker} \citep{Price-Whelan2017} to investigate orbital parameters using the RV measurements for each of the four nights of observation, as listed in Table \ref{table: rvs}. \texttt{The Joker} is a Monte Carlo sampler that produces a converged sample of Keplerian orbital parameters for sparse radial velocity data.

We defined broad priors in our parameter search. By default, most priors in \texttt{The Joker} are set as Gaussian distributions centered on zero where the standard deviation is the tunable parameter. We set a uniform prior on the orbital period, $P_{orb}$, between 1 and 1000 days. \texttt{The Joker} imposes a Gaussian prior on the velocity semi-amplitude, $K$, and $v_{0}$, the barycenter or systemic velocity. By default, both Gaussian distributions are centered at 0 km s$^{-1}$. We set the standard deviation on the prior for $K$ at 200 km s$^{-1}$. We changed the mean of the Gaussian distribution for $v_{0}$ to -214 km s$^{-1}$, the local gas velocity as measured by \citet{Koch2018} and set the standard deviation on the prior for $v_{0}$ at 100 km s$^{-1}$ to keep the distribution fairly broad. Once the priors are set, \texttt{The Joker} generates a user-defined number of prior samples that will then be passed through \texttt{The Joker}'s rejection sampler to create a down-sampled set of posterior samples. We generated 250,000 prior samples and then used the rejection sampler to create a total number of 256 posterior samples.

We then used a standard Markov Chain Monte Carlo (MCMC) sampler \texttt{pymc3} \citep{pymc2023} to fully explore the posterior probability distribution function (PDF). We used the same prior that was used for \texttt{The Joker}'s rejection sampler. The resulting posterior PDF for $P_{orb}$ had two peaks, one at $\sim$2.7 days and one at $\sim$5.3 days with the stronger peak at $\sim$5.3 days. Both of these peaks lie within the 3$-$60 day orbital period range that is typical for supergiant HMXBs \citep{FornasiniAntoniou2023}. The posterior PDF for eccentricity was broad with values ranging from 0 to 0.8, with the highest peak close to 0. The posterior PDF for the velocity semi-amplitude, $K$, was fairly well constrained with a single peak at 15.5$^{+19.1}_{-8.5}$ km s$^{-1}$. The posterior PDF for the systemic velocity, $v_{0}$, was also well constrained with a single peak at -293.6$^{+10.0}_{-7.3}$ km s$^{-1}$. 

Given the sparse sampling of our radial velocity measurements, these orbital parameters should be considered more as a guide for future observations, which are needed to provide the time resolution necessary to more fully sample the system's orbital period and definitively constrain the orbital elements, including orbital period. However, we explore the implication of the results presented in this section in Section \ref{sec:co_mass}.

\section{Discussion}\label{sec:discussion}
In this section we discuss the implications of the measured Balmer line diagnostics (EW, FWHM, line asymmetry), constraints on the system velocity and formation location, and constraints on the compact object mass.

\subsection{Discussion of Balmer Lines}
We observe time evolution in the Balmer line profiles, particularly the shape of the H$\alpha$ line. To understand the evolution of the Balmer lines, we look to previous studies investigating optical emission lines in other HMXB systems with O-type donor stars. Time series spectroscopic observations of O-type donor stars at optical wavelengths are exceedingly rare, but one of the most well-studied systems is Cygnus X-1. Cygnus X-1 is a wind-fed HMXB system comprised of an O-giant donor star and a black hole with masses of $M_{donor}=40.6^{+7.7}_{-7.1}$ $M_{\odot}$ and $M_{BH}=21.2 \pm2.2$ $M_{\odot}$, respectively \citep{Miller-Jones2021}.

Various properties of the H$\alpha$ line including its EW and line profile shape are expected to change with both orbital phase and X-ray spectral state \citep{Gies2003,Yan2008,Brigitte2025}. In Cygnus X-1, the H$\alpha$ line includes components from both the stellar atmosphere and a focused wind component, which is located between the donor star and the compact object within the system \citep[see Figure 2 in][for an illustration]{Brigitte2025}. \citet{Brigitte2025} performed simultaneous X-ray and optical observations of Cygnus X-1 and observed the time evolution in its H$\alpha$ and He I and He II features with orbital phase and X-ray spectral state. They disentangled the contribution from the stellar photosphere, focused wind, and Earth's atmosphere to the H$\alpha$ line and some of the He I and He II lines. They found that the He I lines only show absorption features in the stellar atmosphere component and were not detected in the focused wind while the observed H$\alpha$ line has both a stellar atmosphere and focused wind component. At different points in the system's orbit, the component of the H$\alpha$ line from the stellar photosphere appears in absorption or with a P Cygni profile. The component of the H$\alpha$ line from the focused wind is more strongly blueshifted and is created via photoionization by both the star and the compact object's accretion disk. They did not present a discussion of the H$\beta$ line profile evolution. 

Optical observations of the HMXB system Vela X-1 presented by \citet{Kaper1994} show an increase in blueshifted absorption in the H$\beta$ line, which looks similar in shape to the H$\beta$ line profile we observe in Figure \ref{fig:balmer_line_evolution}. The origin of this feature in Vela X-1 is thought to be a photo-ionization wake caused by the passage of the compact object in its orbit around the O-giant donor star \citep{Kaper1994,Kretschmar2021}.

We observe variable H$\alpha$ emission from the star presented in this paper. It has an asymmetric line profile with and enhanced blue wing at each epoch, its centroid velocity is offset from both the local HI gas and the He I radial velocity, and its EW and FWHM change across epochs. This variable H$\alpha$ emission could be explained with a two component line origin coming from both the stellar photosphere and from a focused wind component. The H$\beta$ line profile that we observe is asymmetrical with enhanced blueshifted absorption. The line is not as variable in shape as the H$\alpha$ line with fairly consistent EW and FWHM across epochs. The centroid velocity of the H$\beta$ line suggests that it likely originates at the stellar photosphere, but the enhanced blue wing of the absorption line may be caused by a photoionization wake similar to what is observed in Vela X-1.

M33 X-7 is another well known HMXB system with an O-type donor star \citep{Orosz2007,Ramachandran2022}. While M33 X-7 has been well studied, optical spectroscopy of the donor star typically focuses on bluer wavelengths, and there is no published analysis of the H$\alpha$ line profile evolution with orbital phase in the literature.

\subsection{System Velocity and Origin}\label{sec:system_velocity}
We use the spatially resolved recent star formation history (SFH) maps of M33 \citep{Lazzarini2022}, a catalog of young star clusters in M33 \citep{Johnson2022,Wainer2022},  and the system velocity estimated in Section \ref{sec:the_joker} to investigate the system's origin. The SFH maps and cluster catalog were both produced using data from the PHATTER survey \citep{Williams2021}.

We assume that the current mass of the candidate donor star measured with SED fitting is $\sim75$ $M_{\odot}$, although we must assume that its initial mass was lower. The evolution of HMXB systems involves significant mass transfer prior to the supernova of the primary star \citep{VanDenHeuvel2019}. Theoretical modeling of M33 X-7 suggests that for a donor star with a current mass of $\sim$70 $M_{\odot}$, the total mass transferred from the primary onto the secondary prior to the primary's supernova could be approximately 35$-$40 $M_{\odot}$ \citep{Valsecchi2010}. Assuming approximately the same amount of material was transferred onto the 75 $M_{\odot}$ secondary in this system, its initial mass would have been closer to $\sim$35 $M_{\odot}$. When investigating potential birthplaces for the candidate HMXB system, we thus assume that the site of formation must have produced at least two stars with $M>$35 $M_{\odot}$ because the primary star, which would now be the compact object in the HMXB system,  was likely initially the more massive star in the system.

\begin{figure*}
\centering
\includegraphics[width=0.9\textwidth]{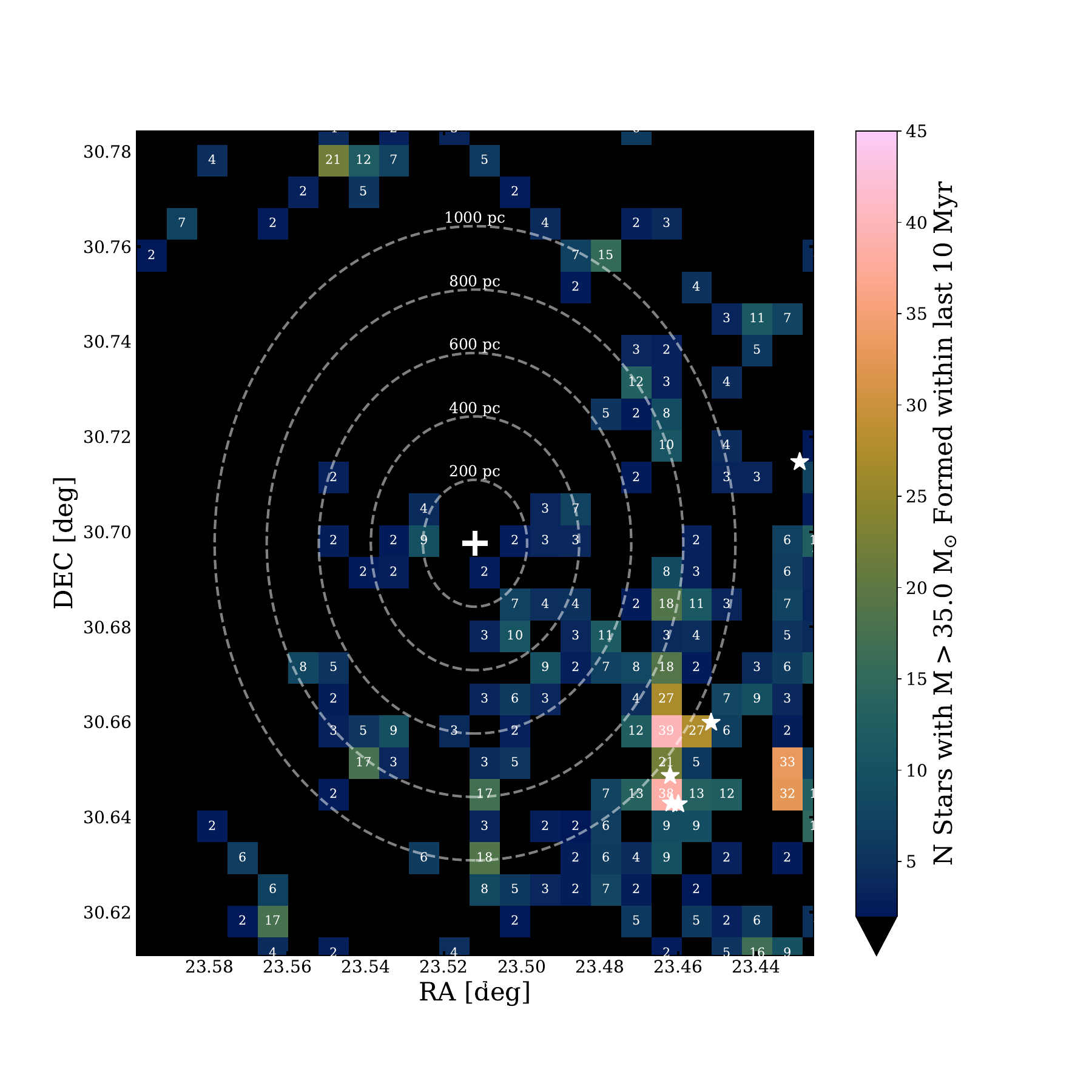}
\caption{Map showing the number of stars with $M>$35 $M_{\odot}$ that are expected to have formed in the last 10 Myr using the SFH measured by \citet{Lazzarini2022}. The ``+'' sign at the center of the figure marks the location of the candidate HMXB donor star discussed in this paper. Radii of 200, 400, 600, 800, and 1000 pc from the HMXB donor star's location are shown with white dashed lines. Each background square indicates a region in the SFH maps. The number of stars with $M>$35 $M_{\odot}$ that would have formed within the last 10 Myr is shown in white text plotted on each region with at least two. Stars mark the location of clusters \citep{Johnson2022,Wainer2022} with ages of $<$10 Myr and $>$90\% probability of having formed at least two stars with $M>$35 $M_{\odot}$ stars based on its current mass.}
\label{fig:sfh_map_age}
\end{figure*}

The spatially resolved recent SFH maps from \citet{Lazzarini2022} divide the PHATTER survey area into spatial regions that measure $\sim$100 pc on a side. Based on these maps, we calculate the number of stars with $M>$35 $M_{\odot}$ formed within the last 10 Myr in the regions surrounding the candidate HMXB system, shown in Figure \ref{fig:sfh_map_age}. This adopted timesecale is a conservative estimate of the maximum lifetime of the secondary O-star. We plot the position of the candidate donor star, with dashed lines indicating distances of 200$-$1000 pc from its current position. Since the mass of the primary is assumed to be larger than the mass of the candidate donor star, it would need to have formed in a region that has a SFH capable of having formed at least two stars with $M>$35 $M_{\odot}$ within the last 10 Myr.

In Figure \ref{fig:sfh_map_age}, there are a few regions that formed more than $\sim$20 or 30 stars with $M>$35 $M_{\odot}$ in the lower right hand of the plot (R.A. $\sim$23.46, Dec. $\sim$30.66) located between $\sim$800 and $\sim$1000 pc from the star's current location. While there are a few very nearby regions (within 200$-$400 pc of the HMXB candidate) that formed 7$-$10 such stars, the sheer number of massive stars formed in this region makes it the most likely origin for the candidate HMXB. The HMXB candidate's current distance from this region of high star formation implies a transverse velocity of about 100 km s$^{-1}$, assuming it has traveled $\sim$1000 pc in the last 10 Myr.

In addition to using the SFH maps to constrain the HMXB system's potential origin, we used the young star cluster catalog for M33 created with the PHATTER data by \citet{Johnson2022,Wainer2022} to identify clusters that could be the system's potential birthplace. We selected clusters with ages $<$10 Myr, which is our conservative upper end on the age of our candidate HMXB system. We then selected clusters with masses that would suggest they have at least a 90\% probability of having produced two stars with $M>$35 $M_{\odot}$, assuming a Poisson sampling of the Kroupa initial mass function \citep{Kroupa2001}, which we find to be clusters with masses $\gtrsim3\times10^{3}$ $M_{\odot}$. We plot the location of these clusters in Figure \ref{fig:sfh_map_age} as white stars. There are five clusters within $\sim$1.2 kpc of the HMXB candidate that meet these criteria. Four of these clusters are within the high star formation region identified from the SFH analysis. The presence of these massive, young clusters paired with the high recent star formation mapped in the SFH both suggest that this is a likely birthplace of our candidate HMXB system.

We can use the star's transverse velocity to determine if it is moving fast enough, $\sim$100 km s$^{-1}$, to have originated from the potential birthplace identified with the M33 SFH maps and cluster catalog. While we cannot measure the star's transverse velocity from its spectrum, we can estimate its order of magnitude using the star's measured radial velocity. In Section \ref{sec:the_joker} we measure a systemic velocity for the system of $\sim$-293 km s$^{-1}$ based on its radial velocity measured during each of the four observing epochs. Compared to the local HI gas velocity at the location of the star, $\sim$-214 km s$^{-1}$ \citep{Koch2018}, this gives a peculiar radial velocity of $\sim$80 km s$^{-1}$. If we assume that this is one component of its 3-dimensional peculiar velocity, and we assume that the other two components are of similar magnitude, that would imply a 2-dimensional transverse velocity of $\sim$ 100 km s$^{-1}$. This transverse velocity is slightly higher than, but broadly consistent with, the range of observed 3-dimensional peculiar velocities ($\sim$5$-$90 km s$^{-1}$) measured in Galactic HMXBs with Gaia data \citep{Fortin2022}.

\subsection{Constraints on Compact Object Mass}\label{sec:co_mass}
As discussed in Section \ref{sec:results}, the orbital parameters fit with \texttt{The Joker} should be treated as preliminary constraints. However, it is instructive to know if these orbital parameters would be consistent with an HMXB. We can use the best fit values of $P_{orb}$ and $K$ to calculate the best-fit binary mass function, $f(M)$, defined below. 
\begin{equation}
f(M)=\frac{P_{orb}K^{3}}{2\pi G}=\frac{M_{2}^{3}\sin^{3}i}{(M_{1}+M_{2})^{2}}
\end{equation}

In this equation, $P_{orb}$ is the orbital period of the binary, $K$ is the velocity semi-amplitude, $M_{2}$ is the mass of the unseen companion, $M_{1}$ is the mass of the donor star for which we can measure an RV, and $i$ is the inclination of the system. Using the best fit values for $P_{orb}$ and $K$ from \texttt{The Joker}, we can determine a best-fit value for $f(M)$ and calculate the minimum mass of the compact object in the HMXB system, $M_{2}\sin(i)$. Since the posterior PDF for $P_{orb}$ was double peaked, we perform the calculation with both peaks: $\sim$2.7 days and $\sim$5.3 days. The velocity semi-amplitude, $K$, returned a single-peaked posterior PDF and a best fit value of 15.5$^{+19.1}_{-8.5}$ km s$^{-1}$. We assume 0 eccentricity and the mass of the donor star, $M_{1}$, to be the current mass measured with SED fitting, $\sim$75 $M_{\odot}$. We can infer a minimum mass for the unseen compact companion of $\sim$1.8 $M_{\odot}$ and $\sim$2.3 $M_{\odot}$ for the shorter and longer orbital period, respectively. As discussed in Section \ref{sec: sed_fitting}, SED fitting may have overestimated the mass of the donor star, which would affect the implied compact object mass. For example, reducing the mass of the donor, $M_{1}$ to 50 $M_{\odot}$ drops the inferred mass of the compact object to $\sim$1.4 $M_{\odot}$ and $\sim$1.8 $M_{\odot}$ assuming the orbital periods above, respectively. We do not know the system's inclination, so the minimum mass measured with both the shorter and longer orbital periods is consistent with both a neutron star, or black hole companion. Additional spectroscopic observations with higher time resolution are needed to more fully sample the system's orbital period and more confidently constrain the orbital elements and the compact object's minimum mass.

\section{Conclusions}\label{sec:conclusions}
We present four epochs of optical spectroscopy for a candidate HMXB donor star in M33. Our main conclusions are as follows:
\begin{enumerate}
    \item The star's optical spectrum confirms that it is an O-type star, although higher resolution spectroscopy that extends to bluer wavelengths is needed to further constrain the spectral type.
    \item We measure the star's radial velocity using the He I $\lambda$6679~\AA~ line and observe a radial velocity shift over the course of the approximately 1 year baseline of our four observations. 
    \item We investigated the system's orbital elements using \texttt{The Joker} and the measured He I radial velocities. The system has a systemic velocity of $\sim$ 80 km s$^{-1}$ relative to the local HI gas and a velocity semi-amplitude of $\sim$15 km s$^{-1}$. The posterior PDF for the system's orbital period is double peaked with one peak at $\sim$2.7 days and one peak at $\sim$5.3 days. The best-fit orbital period and velocity semi-amplitude can be used to calculate the minimum mass of the unseen compact companion, assuming a 75 $M_{\odot}$ donor star as measured with SED fitting of its HST photometry. The measured minimum mass of the unseen companion is consistent with either a neutron star or black hole.
    \item The line profile shape and offset velocity of the H$\alpha$ line suggest it originates from a focused wind rather than the star's surface, which is consistent with observations in other HMXB systems with O-star donors.
    \item The system has a peculiar radial velocity of $\sim$80 km s$^{-1}$ based on constraints of the systemic velocity from radial velocity measurements and the local HI gas velocity. Using SFH measurements and the ages and masses of clusters in M33, we suggest that a plausible origin for the system is a region located about 1000 pc away, which would suggest a transverse velocity of $\sim$100 km s$^{-1}$, which is consistent with the observed radial velocity.
    \item Additional observations are required to fully characterize this system. Higher resolution spectroscopy extending to bluer/UV wavelengths will provide more detailed constraints on the donor star's temperature, mass, and stellar winds. Higher resolution spectroscopy will also allow reduced errors on the RV measurements, which is needed to constrain the system's orbital parameters given the low amplitude of the RV variations.  Additional epochs of spectroscopy with higher time resolution are required to fully sample the system's orbit and place definitive constraints on the system's orbital elements.
\end{enumerate}

\begin{acknowledgements}
Margaret Lazzarini was supported by an NSF Launching Early-Career Academic Pathways in the Mathematical and Physical Sciences (LEAPS-MPS) Award AST-2418745. The work of Daniel Stern was carried out at the Jet Propulsion Laboratory, California Institute of Technology, under a contract with the National Aeronautics and Space Administration (80NM0018D0004).
\end{acknowledgements}

\software{Astropy \citep{astropy:2013, astropy:2018, astropy:2022}, Matplotlib \citep{Hunter:2007}, BEAST \citep{Gordon2016}, The Joker \citep{Price-Whelan2017}, lmfit \citep{lmfit}, Crameri Scientific Colour Maps \citep{crameri_software,Crameri2020}, PyMC \citep{pymc2023}}

\bibliography{references}
\bibliographystyle{aasjournalv7}
\end{document}